# Band gap tuning in ferroelectric $Bi_4Ti_3O_{12}$ by alloying La$TM$O$_3$ ($TM$ = Ti, V, Cr, Mn, Co, Ni, and Al)

Woo Seok Choi and Ho Nyung Lee

*Materials Science and Technology Division, Oak Ridge National Laboratory, Oak Ridge, TN 37831, USA*

**We fabricated ferroelectric $Bi_4Ti_3O_{12}$ (BiT) single crystalline thin films site-specifically substituted with La$TM$O$_3$ ($TM$ = Al, Ti, V, Cr, Mn, Co, and Ni) on SrTiO$_3$ substrates by pulsed laser epitaxy. When transition metals are incorporated into a certain site of the BiT, some of BiT-La$TM$O$_3$ showed a substantially decreased band gap, coming from the additional optical transition between oxygen 2$p$ and $TM$ 3$d$ states. Specifically, all alloys with Mott insulators revealed a possibility of band gap reduction. Among them, BiT-LaCoO$_3$ showed the largest band gap reduction by ~1 eV, positioning itself as a promising material for highly efficient opto-electronic devices.**

Ferroelectric materials are reemerging as promising candidates for opto-electronic device applications.[1] By utilizing the spontaneous polarization, which can induce internal electric field, efficient photoelectric response could have been observed.[2-4] One of the major obstacles that limit the photoelectric application of ferroelectrics is their wide band gap. Since most of ferroelectrics have wide band gaps of above 3.3 eV, absorbing light in the visible wavelength range is rather challenging.[5] Therefore, identifying approaches to reducing the band gap of ferroelectrics without losing the useful ferroelectricity would bring significant scientific and technological breakthroughs with complex metal oxides.

Recently, we showed that alloying $LaCoO_3$ with ferroelectric $Bi_4Ti_3O_{12}$ (BiT) can be one of the approaches to systematically lower the band gap of a ferroelectric.[6] Particularly, site specific substitution of layered ferroelectric $Bi_4Ti_3O_{12}$ (BiT) with La and Co could considerably reduce the band gap of BiT by ~1 eV, while maintaining strong ferroelectricity. From density functional theory calculations, the decrease of the band gap was attributed to an additional Co electronic state just below the conduction band of BiT. In this study, La based perovskite transition metal oxide was chosen as alloying material since La doped BiT is an excellent ferroelectric.[7,8] Among other $LaTMO_3$ ($TM$ = Al and 3$d$ transition metals), $LaCoO_3$ was chosen because it had reasonably low band gap and the largest absorption in the visible wavelength range. Although $LaCoO_3$ might be the best candidate to reduce the band gap of BiT as it has a large absorption coefficient, it is worthwhile to examine other $LaTMO_3$ in order to see if the band gap of BiT can further be modified with other $TM$ substitution. This investigation could help attain a deeper understanding on the band gap reduction mechanism in BiT.

In this work, various $LaTMO_3$ ($TM$ = Al, Ti, V, Cr, Mn, Co, and Ni) were alloyed with BiT and their structural and optical properties were examined. X-ray diffraction study showed that most of BiT-$LaTMO_3$ preserved the characteristic layered crystal structure of BiT, suggesting site-specific substitution is valid for most BiT-$LaTMO_3$. We could observe a reduced band gap for $TM$ = V, Cr, Mn, Co, and Ni, resulting from an additional electronic state due to the *TM*. The largest band gap change was observed for BiT-LaCoO$_3$, indicating that Co is the key element in reducing the band gap of the ferroelectric BiT.

We used pulsed laser epitaxy to fabricate (001)-oriented BiT-La*TM*O$_3$ epitaxial films on TiO$_2$-terminated single crystal (001) SrTiO$_3$ (STO) substrates. Samples were grown at 700 $^0$C in 100 mTorr at the repetition rate of 10 Hz. Half unit cell layers of La*TM*O$_3$ and quarter unit cell layers of BiT were alternatingly ablated from stoichiometric La*TM*O$_3$ and BiT targets. More detailed growth method is described elsewhere.[6] It should be noted that a similar alloying study has been recently reported for LaFeO$_3$-Bi$_4$Ti$_3$O$_{12}$, which focused on its magnetic properties.[9] While they used similar pulsed laser deposition, they used a single composite target to fabricate thin films, instead of using a controlled growth with two different targets. This might have resulted in additional layer of LaFeO$_3$ in their study, instead of site-specific substitution as in our case. However, we note that the crystal quality is considerably enhanced using the approach described in this paper, judging from both x-ray diffraction (XRD) and Z-contrast transmission electron microscopy.[6,9] The crystal structure of films was characterized using XRD (X'Pert, Panalytical Inc.). Spectroscopic ellipsometry (M-2000, J. A. Woollam Co.) was used to obtain optical conductivity spectra ($\sigma_1(\omega)$) of the thin films between 1.5 and 5 eV at room temperature.[10]

Schematic drawings of crystal structures of BiT, La*TM*O$_3$, and site-specifically substituted BiT-La*TM*O$_3$ are shown in Fig. 1. BiT belongs to the Aurivillius family with a highly anisotropic pseudo-orthorhombic unit cell ($a$ = 5.450, $b$ = 5.406, and $c$ = 32.832 Å).[11] There are two distinct alternating layers. One fluorite-like layer (Bi$_2$O$_2$, Bi shown in yellow) and two perovskite layers (Bi$_2$Ti$_3$O$_8$, Bi shown in blue) are connected with an additional oxygen plane. On the other hand, La*TM*O$_3$ has simple perovskite structure, where the pseudo-cubic lattice parameter ranges between 3.80 and 3.93 Å, which offers a good lattice match with BiT and also the STO substrate. When BiT and one of the La*TM*O$_3$ (particularly *TM* = Co) are alloyed together, it has been shown that La specifically substitutes upper Bi in Bi$_2$O$_2$ layer. Moreover, it has been suggested that Co substitutes Ti near the Bi$_2$O$_2$ layer, forming the self-ordered BiT-La*TM*O$_3$ superstructure shown in the right side of Fig. 1.[6] Since Co and other $TM$s studied here have similar ionic properties, *e.g.*, ionic radius, electronegativity, etc., similar site-specific substitution can be expected for most BiT-La$TM$O$_3$.

Figure 2 shows x-ray $\theta$-$2\theta$ diffraction patterns for BiT and BiT-La$TM$O$_3$ films deposited on STO substrates. Note that BiT has about eight times longer *c*-axis unit cell lattice constant compared to simple perovskites. Therefore, the four peaks (002, 004, 006, and 008) in the same $2\theta$ angle range where STO substrate shows one peak (denoted as star in Fig. 2) clearly confirm the characteristic layered structure. We further note that the layered feature of BiT is more or less well preserved even after alloying with various La$TM$O$_3$. While the lowest index (002) peak intensity decreases significantly for some of the BiT-La$TM$O$_3$ (*e.g.* $TM$ = Al, V, and Ni) possibly due to the induced disorder, other peaks are clearly present and only show little shift in $2\theta$. This indicates that the overall layered structure of BiT is rather robust, *i.e.*, La$TM$O$_3$ is alloyed into BiT by substitution, and is not forming an additional epitaxial layer, as schematically shown in Fig. 1. In addition, the absence of other structures in XRD, except for the film and substrate peaks ensures that there are no impurity phases in the films. It should be noted that the peaks for $TM$ = V and Cr are rather broad compared to other $TM$s. This might be due to the large ionic radius of V and Cr ions. When large ions like V or Cr substitutes for Ti, they might need more rooms for adjustments. Gradual enhancement in XRD peak sharpness as increasing the atomic number, *i.e.* from V to Ni, seems to support the scenario.

Spectroscopic ellipsometry confirms that the band gap of BiT can be modified by site-specific substitution with La$TM$O$_3$, at least for $TM$ = V, Cr, Mn, Co, and Ni. Figure 3 shows $\sigma_1(\omega)$ from BiT and BiT-La$TM$O$_3$ films. For BiT, strong optical absorption starts at ~3.55 eV, manifesting the charge-transfer gap of the pure material. Two distinct peak features are shown at ~3.8 and ~4.7 eV, originating from the transition between O 2$p$ and Ti 3$d$ $t_{2g}$ and $e_g$ states, respectively.[12] For BiT-La$TM$O$_3$ with $TM$ = Al and Ti, we could not observe any band gap change, although the peak shapes are somewhat different. This indicates that La incorporation does not affect the band gap of BiT, clearly stressing the importance of *TM* in tuning the band gap. Moreover, non-transition-metal Al also does not change the band gap. It should further be noted that doping with non-transition-metal-elements such as Ga, In, or Nd is also known to be an ineffective method to change the band gap of BiT significantly.[13,14]

On the other hand, when transition metals (*TM* = V, Cr, Mn, Co, and Ni) with *d* electrons are used for the substitution, BiT-La*TM*O$_3$ shows additional absorption in $\sigma_1(\omega)$, denoted as triangles in Fig. 3. This additional absorption is located at lower photon energy, typically between 3.5 and 3.6 eV, reducing the band gap of BiT-La*TM*O$_3$. For Co, it has been previously suggested that this additional state is coming from the transition from O 2*p* to Co 3*d* state.[6] Similar analogy can be applied to other transition metals, judging from the similarity in the energy scale and absorption peak intensity. Presence of 3*d* electron from various transition metals forms 3*d* electronic states within the gap of BiT, decreasing the band gap of the alloyed material.

Figure 4 summarizes the band gap of BiT and BiT-La*TM*O$_3$ obtained from $\sigma_1(\omega)$. Values of different types of band gaps (charge transfer and Mott gaps) for La*TM*O$_3$ are extracted from the literatures (LaTiO$_3$, LaVO$_3$, LaCrO$_3$, LaMnO$_3$, LaCoO$_3$, and LaNiO$_3$).[15-22] and are also shown for comparison. The band gap value of BiT-La*TM*O$_3$ seems to follow the trend of charge transfer gap value, which is reduced with increasing the number of *d* electrons in $TM^{3+}$. This experimental trend is as expected, since the additional state responsible for the band gap reduction is mainly coming from the transition from O 2*p* to *TM* 3*d* state, which constitutes the charge transfer gap in La*TM*O$_3$. One exception is Ni, where the band gap of BiT-LaNiO$_3$ increases above the band gap of BiT-LaCoO$_3$. This might be related to the fact that LaNiO$_3$ is a metal, while other La*TM*O$_3$ studied here are all insulators. Therefore, we can experimentally conclude that cobalt is the best material to reduce the band gap of BiT through site-specific substitution technique. Further studies based on more complicated considerations including oxygen vacancy and other valence state of *TM* ions might shed light on the detailed electronic structure changes in BiT-La*TM*O$_3$.

In summary, we experimentally investigated the structural and optical properties, and band gap change in La$TM$O$_3$-substituted ferroelectric Bi$_4$Ti$_3$O$_{12}$, with $TM$ = Al, Ti, V, Cr, Mn, Co, and Ni. When Ti in Bi$_4$Ti$_3$O$_{12}$ is substituted with another transition metal element, the resulting new composite shows an additional absorption below the band gap of Bi$_4$Ti$_3$O$_{12}$ yielding in a decreased band gap. In particular, Bi$_4$Ti$_3$O$_{12}$-LaCoO$_3$ shows the largest band gap change (~1 eV reduction) among the transition metals substitutes.

**ACKNOWLEDGMENT**

We appreciate useful discussion with Khuong P. Ong. This work was supported by the U.S. Department of Energy, Basic Energy Sciences, Materials Sciences and Engineering Division.

**FIGURE CAPTIONS**

FIG. 1. (Color online) Schematic diagrams of BiT, La$TM$O$_3$, and BiT-La$TM$O$_3$. A half pseudo-orthorhombic unit cell of BiT and one unit cell of La$TM$O$_3$ are schematically shown. When they are alloyed together, specific sites in BiT are specifically substituted with La and $TM$, resulting in a new artificial material shown as BiT-La$TM$O$_3$.

FIG. 2. (Color online) XRD $\theta$-$2\theta$ patterns for pure BiT and BiT-La$TM$O$_3$ films with $TM$ = Al, Ti, V, Cr, Mn, Co, and Ni on STO substrates. The overall Aurivillius structure is clearly conserved, showing most of the even order peaks that correspond to four times of the perovskite unit cell. Peaks from STO substrate are denoted as asterisks.

FIG. 3. (Color online) $\sigma_1(\omega)$ for pure BiT and BiT-La$TM$O$_3$ with $TM$ = Al, Ti, V, Cr, Mn, Co, and Ni obtained from spectroscopic ellipsometry. Below the band gap of BiT, there exists an additional absorption for $TM$ = V, Cr, Mn, Co, and Ni, marked as triangles. Consequent band gap value of each compound is estimated from the onset of the absorption. Vertical offsets are used for comparison of spectra.

FIG. 4. (Color online) Summary of band gaps (red circles) of pure BiT and BiT-La$TM$O$_3$ with $TM$ = Al, Ti, V, Cr, Mn, Co, and Ni. Mott (black squares) and charge transfer (blue triangles) gaps of pure La$TM$O$_3$ are also shown as comparison.[15-22] As a reference, the band gap of pure BiT is indicated with the horizontal thick grey line. BiT-LaCoO$_3$ shows an exceptionally small band gap among the BiT-La$TM$O$_3$ systems tested.

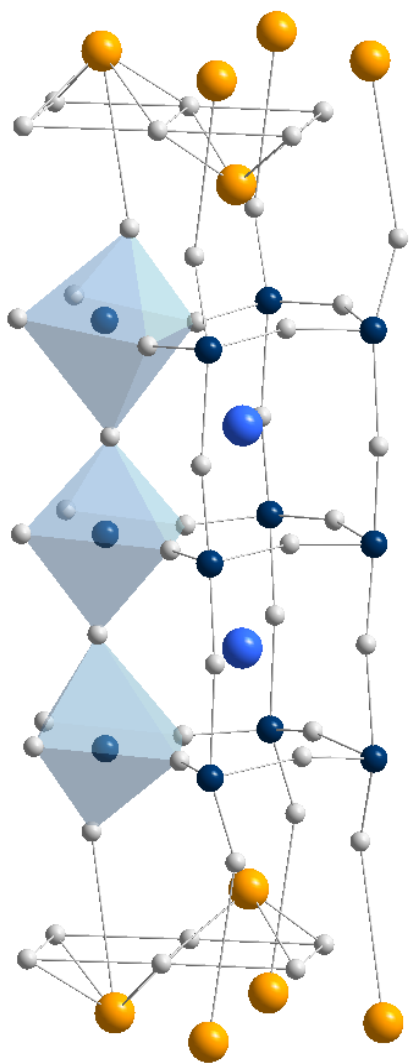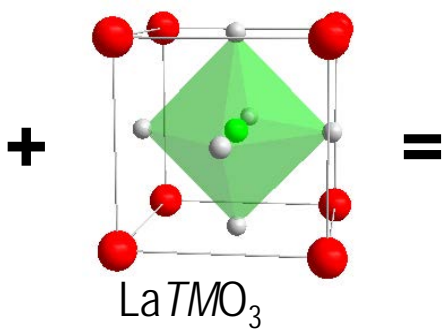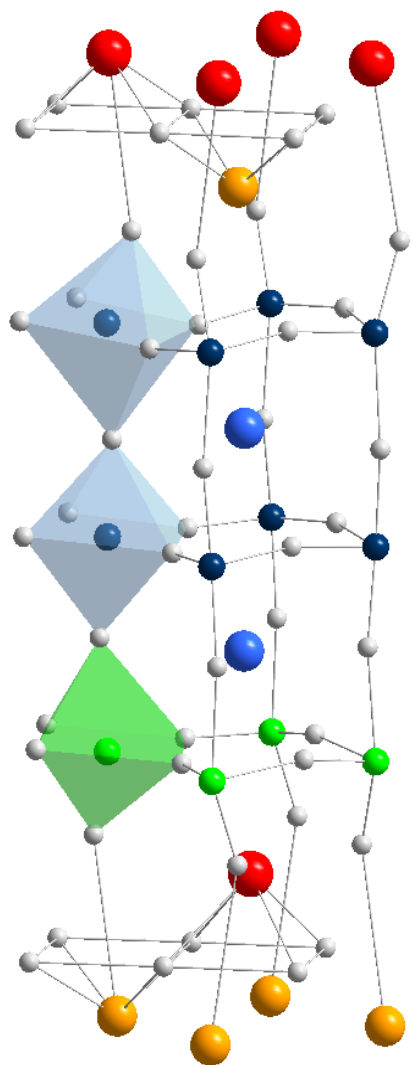

Bi$_4$Ti$_3$O$_{12}$ + La*TM*O$_3$ = Bi$_4$Ti$_3$O$_{12}$-La*TM*O$_3$

- Bi in Bi$_2$O$_2$ plane
- Bi in Bi$_2$Ti$_3$O$_{10}$ plane
- Ti
- La
- *TM*
- O

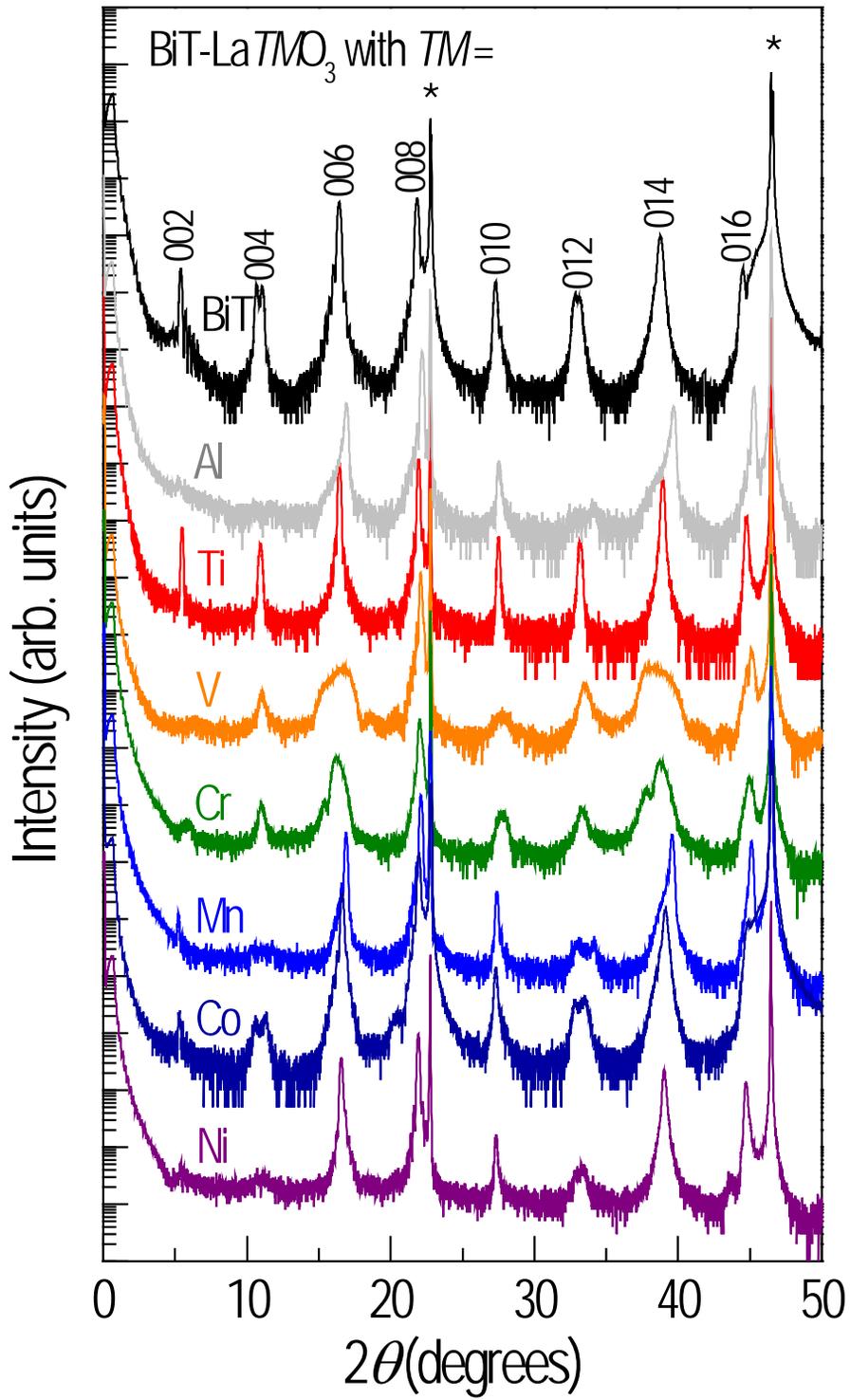

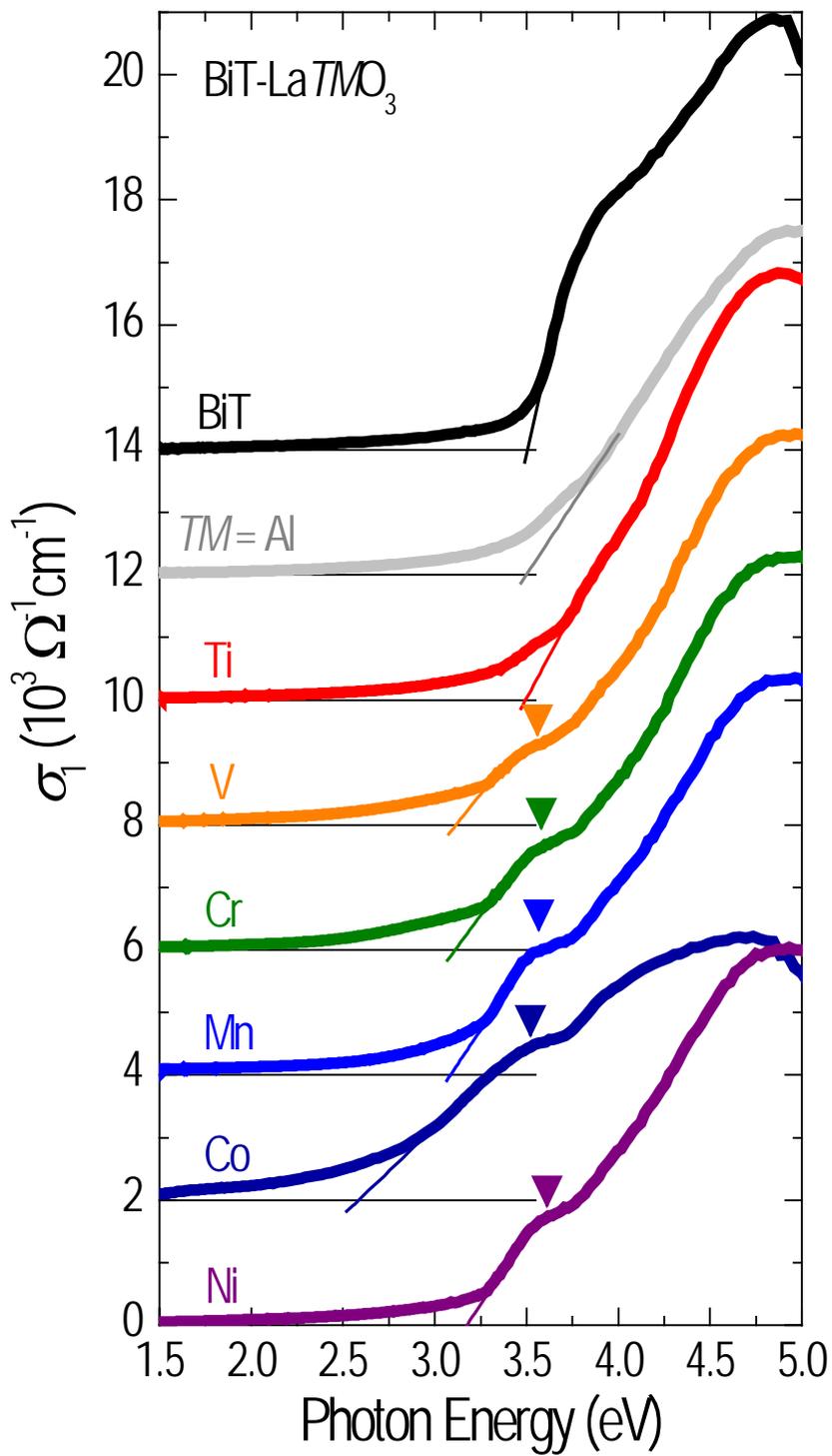

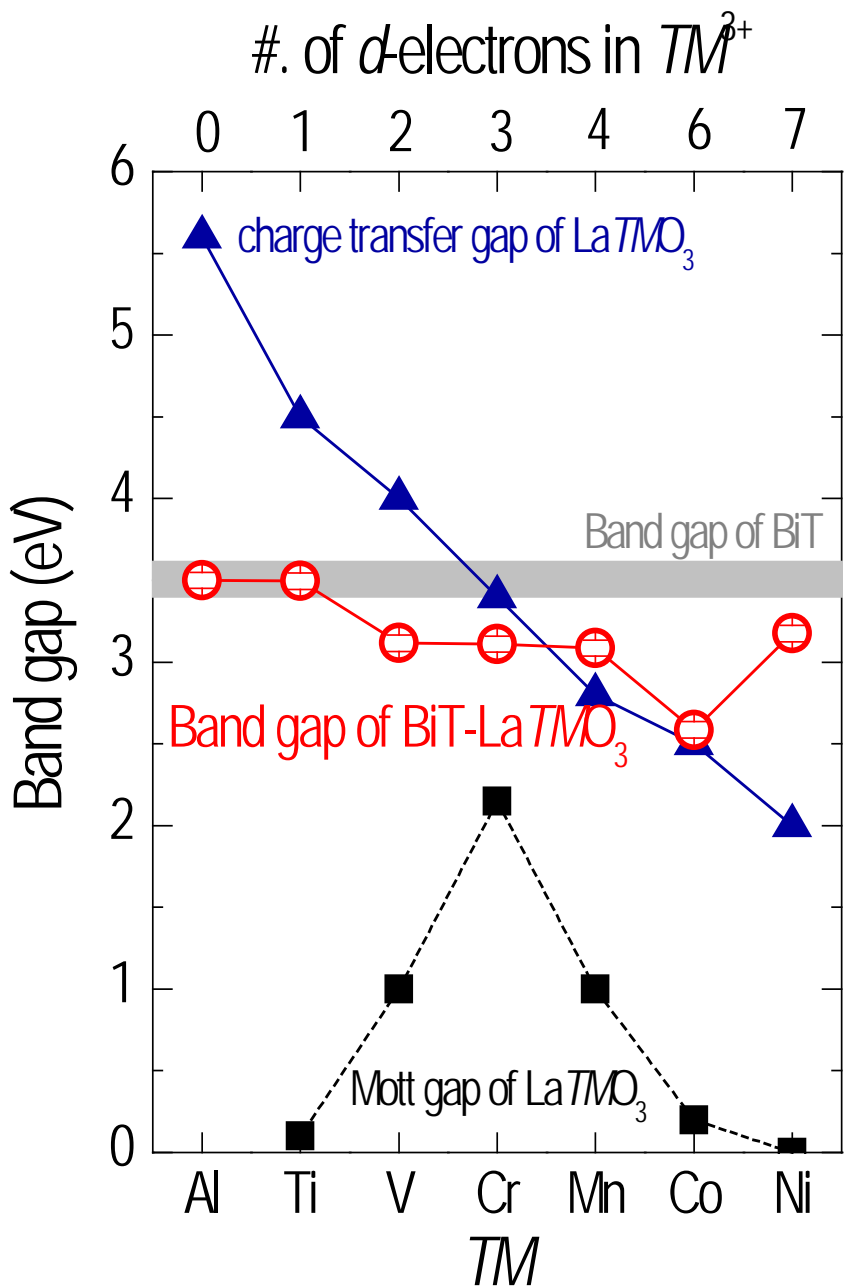